\documentstyle[psfig,epsfig]{article}
\begin{document}
\title{\bf Photoabsorption on nuclei in the shadowing threshold region.}
\author{
 V. Muccifora$^*$, N. Bianchi$^*$, A. Deppman, E. De Sanctis,
 M. Mirazita, E. Polli, P. Rossi.\\
{\em INFN-Laboratori Nazionali di Frascati,}
{\em C.P. 13, I-00044 Frascati, Italy}\\
R. Burgwinkel, J. Hannappel, F. Klein, 
D. Menze, W.J. Schwille, F.Wehnes.\\
{\em Physikalisches Insitut der Universitat Bonn,}
{\em Nussallee 12, D-53155 Bonn, Germany}
} 

\maketitle

\vspace{1cm}
\hyphenation{pa-ram-e-tri-za-tion}
\begin{abstract}
The energy and nuclear mass dependences of the total hadronic cross section 
in the energy range 0.5-2.6 GeV have been measured at Bonn using 
the SAPHIR tagged photon beam.
The measurement, performed on  
C, Al, Cu, Sn and Pb, provides the first photoabsorption data in the region 1.2-1.7 GeV. 
The results  show a significant reduction of the 
photoabsorption strength  on the bound nucleon compared to the free nucleon case
in the whole energy region.
Above 1.2 GeV this reduction  decreases with the nuclear
density and can be interpreted as a signature of a low energy onset of the shadowing effect.
 \end{abstract}
{\bf 

PACS n. 25.20.Gf, 12.40.Vv

}
\vspace{1cm}

{\footnotesize$^*$ Corresponding authors : valeria.muccifora@lnf.infn.it, nicola.bianchi@lnf.infn.it}

\twocolumn

\vfill
\eject

\section{Introduction}

\indent The modifications of the hadron properties and of the elementary couplings 
in the nuclear medium is one of the topics in nuclear
physics currently addressed in various experimental and theoretical investigations.

The properties of baryon resonances in nuclei have been studied  
in recent photoabsorption experiments
at Frascati \cite{BIA93, BIA94, BIA96}, Mainz \cite{MAC96, MAC97}, and  Bonn \cite{MIR97}.
These showed significant
medium effects: while the $\Delta$-resonance is only slightly distorted, higher excited nucleon states N$^*$,
 in the second and third resonance region, are washed out.
Furthermore, for photon energy $k>$ 0.6 GeV the absolute value of the cross section per nucleon is reduced, with
respect to the free-nucleon case.

 The mechanism of this damping is not yet well understood. 
In the resonance region (0.6-1.2 GeV) Fermi-motion and Pauli-blocking alone
 are unable to reproduce the resonance  
disappearance, therefore strong effects in the excitation, propagation and interaction of the 
baryon  \cite{MUC97} and meson  \cite{RAP98, PET98}
resonances in the nuclear medium are advocated.
At higher energies  Vector Meson Dominance (VMD) models predict sizeable 
shadowing effects starting from about 2 GeV \cite{BA78, PI95, BO96}.

 In this paper are reported  the results of the photoabsorption measurements
 on C, Al, Cu, Sn and Pb performed at Bonn between 0.5-2.6 GeV.
The cross sections for C and Pb have been already published \cite{MIR97}.
Here  the cross sections for the other nuclei are given along with the evaluation of the 
 reduction  and the density dependence of the photoabsorption 
strength in different energy regions.

In section 2 the experimental setup and method are extensively described. The analysis procedure 
is reported in section 3.
In section 4  the results of the measurement are presented: specifically the cross sections, the ratio between
photonuclear and photonucleon cross sections, and the nuclear density dependence.

\section{Experimental setup and method}

The photoabsorption measurements were performed using the photohadronic 
method.
This  method consists in measuring the photoproduction yield of
hadronic events with a large solid angle detector, rejecting the vastly preponderant
 electromagnetic events by a separation using a forward angle shower detector. 
 The photohadronic method was successfully applied in all  previous measurements
of the total photoabsorption cross section above the $\Delta$-resonance excitation energy.

\begin{figure}[t]
\vspace{7cm}
\leavevmode
\includegraphics{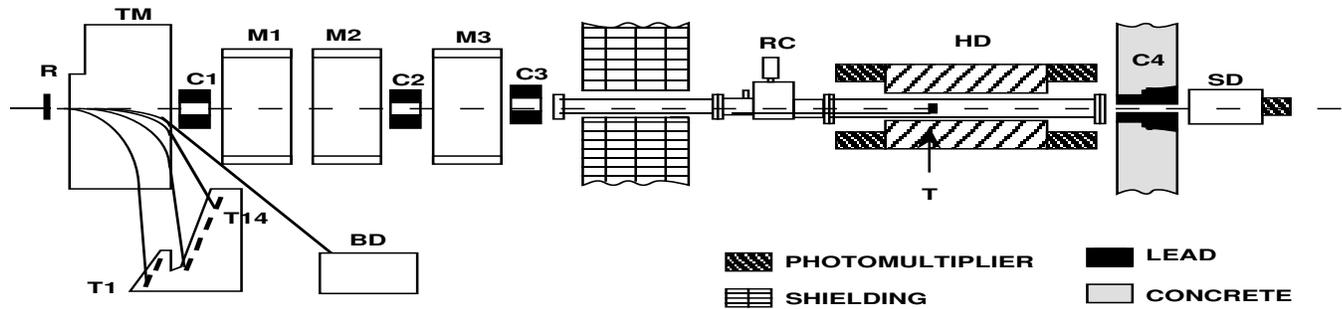}
\caption{Schematic layout side view of the experimental set-up.
(not to scale). R radiator; TM tagging magnet; T1-T14 tagging counters;
BD beam dump; M1-M3 sweeping magnets;
C1-C4 lead collimators; RC remote control system for target movement; T target; 
 HD hadron detector; SD shower detector.}
\label{Fig1}
\end{figure}

The measurement was carried out at Bonn using the SAPHIR tagged photon 
beam \cite{SCH94} of the ELSA accelerator and an apparatus similar to the one previously used at Frascati
for nuclear photoabsorption measurements up to 1.2 GeV\cite{BIA96}.
A schematic layout of the photon beam line and of the detector is given 
in Fig. \ref{Fig1}.

\subsection{Tagged photon beam}

The photon beam was produced by the bremsstrahlung of electrons extracted
from the  ELSA accelerator into the SAPHIR beam line.
The energy of the scattered electrons was measured in the tagging system TOPAS II which
also supplied the trigger for the events.
The tagging system  consisted of a bremsstrahlung radiator, a dipole
magnet and a tagger made by two multiwire proportional chamber (MWPC) placed in front of
scintillation counters. 

As a radiator (R) a copper foil  0.006$X_0$  thick ($X_0$ being the radiation length) has been used.
The dipole magnet  had a maximum deflecting power of 1.2 T$^.$m,
corresponding to a maximum endpoint energy $E_0 = 3.3$ GeV.
 For this measurement, the higher energy-resolution
information from the MWPC of the tagger was not used and the photon energy 
was reconstructed from the timing hodoscope only.
The latter consisted of 14  scintillator counters (T1-T14), each 4.5 cm thick and different in
size, providing a photon energy resolution ranging from about 9\% for the 
lowest energies to about 1\% for the highest energies.
The tagger covered the photon energy range $0.30E_0$-$0.95E_0$.
In this experiment the photon energy
range 0.5-2.6 GeV has been covered, with large overlapping regions
at three electron beam energies $E_0$ = 1.6, 2.2 and 2.8 GeV.

The photon beam passed through a set of collimators and the SAPHIR apparatus.
The three collimators C1-C3 defined a 1.5 cm  diameter  photon beam  at the 
target position, while the three sweeping magnets M1-M3 strongly suppressed the charged background
in the beam.

The lead glass shower detector SD 
allowed the simultaneous measurement of the tagged photon flux for each tagging channel.
The  photon beam intensity and
 the tagging efficiency (defined as the ratio between the number of tagged photons
and the number of counts in the relevant tagging channel)
were measured on-line for each energy interval.

\begin{figure}[ht]
\vspace{7cm}
\leavevmode
\includegraphics{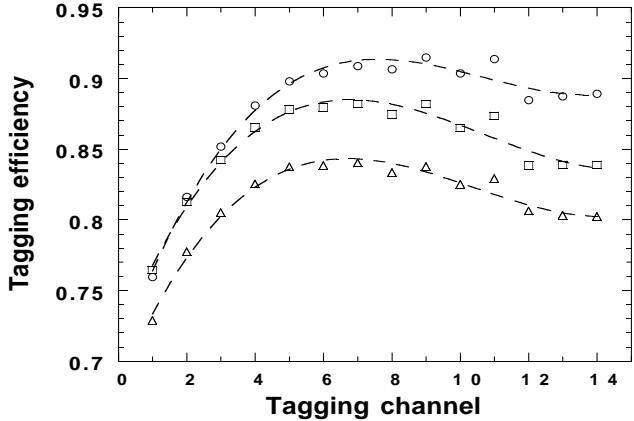}
\caption{Tagging efficiency of the 14  channels for the three different beam energies: 1.6 GeV (triangles), 2.2 GeV (squares)
and 2.8 GeV (circles). Dashed curves are only guides for the eye. }
\label{Fig2}
\end{figure}

In Fig. \ref{Fig2} the  tagging efficiencies  of the
14 tagging channels measured at the three electron beam energies are shown. 
The values range between
 0.73 and 0.92, depending on both the photon and electron beam energies.
The decrease  observed for the first channels (which correspond to the  highest photon energies) is due 
to both the Moller-scattering and the background on the tagging counters.
The decrease observed at the lowest electron beam energy is due
to the wider bremsstrahlung photon emission angle and to the strong collimation cut.

\begin{figure}[ht]
\vspace{7cm}
\leavevmode
\includegraphics{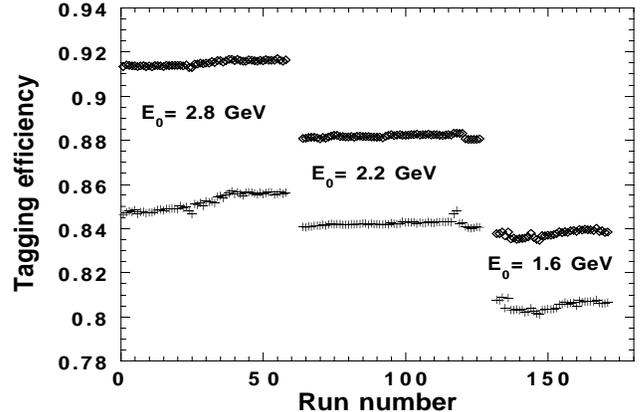}
\caption{Tagging efficiency of  channels T3 (crosses) and T9 (diamonds) measured in each run for the three different
 beam energies.}
\label{Fig3}
\end{figure}

The tagging efficiencies were found stable within $\approx$1\% during the whole data taking,
 as shown 
in Fig. \ref{Fig3} for two sample tagging channels. 

\subsection{Targets}

Solid targets (T) of  C, Al, Cu, Sn and Pb were used. They had the form  
of disks  3 cm in diameter and thicknesses  ranging between 
0.08X$_0$ for C to 0.2X$_0$ for Pb (actual values are given in Table~\ref{Table1}). 

The effective attenuation of the photon beam due to the 
electromagnetic interaction in the target was calculated for each 
nucleus as a function of the photon energy.
The average photon beam attenuation ranged between 3\% for C and 6.5\% for Pb.
The targets were individually mounted on a suitable frame and moved into and off the photon beam by a remote control system (RC).
 
In order to reduce effects due to possible changes in the electron beam, empty frame
 measurements were regularly
 interspersed inside a complete cycle of target runs and their contributions were subtracted. The empty frame yields,
mainly ascribed to the photon beam interactions on materials along the beam line and on the target frame,
were  equivalent to less than one g cm$^{-2}$ of lead and 
 were found stable within $\approx$ 0.8\% during the whole measurement. 

\begin{table}[h,b]
\caption{Target T, thickness [g/cm$^2$], yield (\#) (x$10^4$), and average overall MC corrections
 ($\delta$) at each 
electron beam energy
$E_0$ [GeV]. The empty frame (EF) yield is also given. }
\begin{center}
\begin{tabular}{lcc@{\hspace{0.2cm}}cc@{\hspace{0.2cm}}cc@{\hspace{0.2cm}}c}\hline \hline
T     &Thickness   &\multicolumn{2}{c}{$E_0 = 1.6$}&
                   \multicolumn{2}{c}{$E_0 = 2.2 $}&
                   \multicolumn{2}{c}{$E_0 = 2.8 $} \\
\cline{3-8}
           &                & \#  & $\delta$ & \#  & $\delta$ & \#  & $\delta$ \\ \hline
   C       & 3.450$\pm$0.003 & 22 & 8.4\% & 60 & 5.8\% & 43 & 8.1\% \\
   Al      & 2.399$\pm$0.004 & 20 & 8.0\% & 66 & 5.4\% & 43 & 7.2\% \\
   Cu      & 1.909$\pm$0.013 & 19 & 7.6\% & 62 & 4.3\% & 50 & 6.2\% \\
   Sn      & 1.531$\pm$0.011 & 25 & 5.8\% & 66 & 2.5\% & 53 & 5.0\% \\
   Pb      & 1.240$\pm$0.016 & 35 & 2.7\% & 53 &-0.4\% & 67 & 3.6\% \\
   EF    &                 & 22 &       & 84 &       & 68 &        \\
\hline \hline
\end{tabular}
\label{Table1}
\end{center}
\end{table}

\subsection{Detectors}

A NaI 
crystal hadron detector (HD),  consisting of four cylindrical sectors, 
each 60 cm long and 12 cm thick, surrounding the target, detected the 
charged hadrons and neutral mesons produced by the photon interaction in 
the target.
 The electromagnetically produced leptons and photons, 
 mostly emitted close to the photon beam direction, were vetoed by 
the SD positioned about 1 m downstream to the target. 
Hadronic absorption of a photon of a given energy was indicated by a 
coincidence of signals from the relevant tagging channel and the HD without a 
simultaneous signal in the SD.

\vspace{7cm} 

\vspace*{7cm}
The HD angular coverage was 8$^0 < \theta < 169 ^0$ for the polar angle and almost 2$\pi$ for the azimuthal angle,
which corresponds to more than 98\% of the full solid angle. Due to its thickness,
the  HD  detected about  $ 40\%$ ( $ 30\%$) of the total energy carried  by the
hadrons produced by 0.5 GeV (2 GeV) photons. 
This energy is remarkably higher than that released in the HD by the products from electromagnetic events.

The SD consisted of a dense SF6 lead glass cylinder, 30 cm
 long and 12 cm in diameter. The large detector thickness
 (19X$_0$) provided an efficiency close to unity for detecting  the 
electromagnetic showers generated from the
photon beam and both the Compton photons and the  lepton pairs produced off the target.
The lead collimator C4, placed between the HD and the SD, defined a 
maximum polar angle of 2.4$^0$ with respect to the 
target center.
This 
allowed to detect both the beam photons and the electromagnetic products originating from the target, while
strongly reducing the number of low energy hadrons which might reach the SD. In addition, the threshold of the
\v{C}erenkov process in the SD provided a further rejection of the low energy hadrons.

\subsection{Measurement}

The photon energy range 0.5-2.6 GeV was covered with three different electron beam energies 
 with widely overlapping  photon energy regions.
This allowed a  check of the 
reproducibility of the measurements and gave an estimate of the
systematic errors that could
arise from different running conditions. Detector working parameters, such as HD and SD energy thresholds,
 were adjusted to optimize the efficiency of the hadron detection and to reduce the electromagnetic contamination 
at the different beam energies.
The tagged photon beam rate was kept constant at 5$^.$10$^4$ photons/s in order to reduce
 the random coincidence contamination.
This amounted to about 1\%-6\% of the rate of true events, depending on both
the target and the electron beam energy. Nevertheless, the random coincidences 
were on-line measured and  subtracted.

The number of events collected at the three electron energies, 
for the five targets and for 
the empty frame,
are given in Table~\ref{Table1}.

\section{Procedure and corrections}

The measured hadronic yields were very close to the absolute values of the total cross section, 
the off-line corrections being very small. The latter were due to
 i) the loss of the  events with all hadrons emitted at a polar angle less than the minimum HD detection angle,
 or depositing in the HD an energy below the threshold,
ii) the HD contamination due to the products of  unvetoed electromagnetic events with energy above the HD threshold,   
iii) the SD contamination due to the  events with hadrons releasing energy in both the HD and SD 
above the relevant thresholds.

\begin{figure}[t]
\vspace{7cm}
\leavevmode
\includegraphics{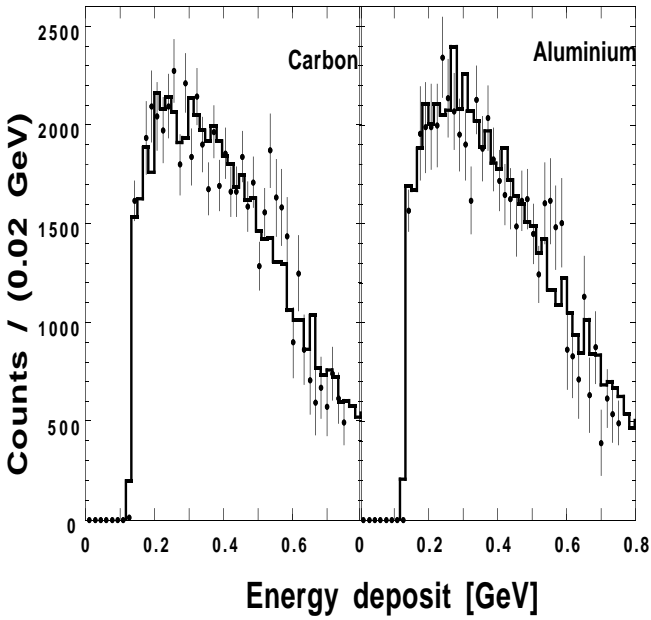}
\caption{Comparison between the simulated (histogram) and measured (solid circles) yield 
of the HD to 0.84-2.66 GeV photons on carbon  and aluminium  targets.
The HD threshold setting was 0.13 GeV.}
\label{Fig4}
\end{figure}

In order to calculate the hadronic corrections i) and iii)
 a Monte Carlo (MC) simulation was developed, based on an
intranuclear cascade model for photonuclear reactions.
This code \cite{IL97} accounts for the photon interaction with nucleons in the target through one-pion, 
two-pions and multi-pions production processes in both resonant and non resonant states; it simulates 
the intranuclear cascade of the photo-hadrons, which leaves the 
residual nucleus in an excited state that emits low-energy evaporation nucleons 
and light nuclei. 
The HD response function to the hadrons generated by this 
 cascade-evaporative code was evaluated by using the Geant-3.21 code.
 Fig. \ref{Fig4} shows the simulated  HD response function to the hadrons photoproduced
on C and Al targets by 0.84$\div$2.66 GeV photons.  
 Also shown in the figure are the measured spectra with a threshold 
cut at 0.13 GeV.
The simulated and measured spectra are in good agreement with each other.
The broad peak  shown at about 0.3 GeV  is due to the hadronic events 
with at least one pion in the final state.

In order to evaluate the correction ii), 
electromagnetic processes were simulated by using a modified version of the Geant-3.21
code, where the experimental energy and angular distributions of pair
production in the energy range of interest were explicitly introduced.
In addition, the \v{C}erenkov photon emission, the attenuation of the \v{C}erenkov light
inside the lead glass and the spectral response of the photomultipliers were taken into account.

Checks of the MC predictions
were performed in order to test the effect of the energy and angular cuts
on the efficiency and the acceptance of HD and SD.

\begin{figure}[t]
\vspace{11cm}
\leavevmode
\includegraphics{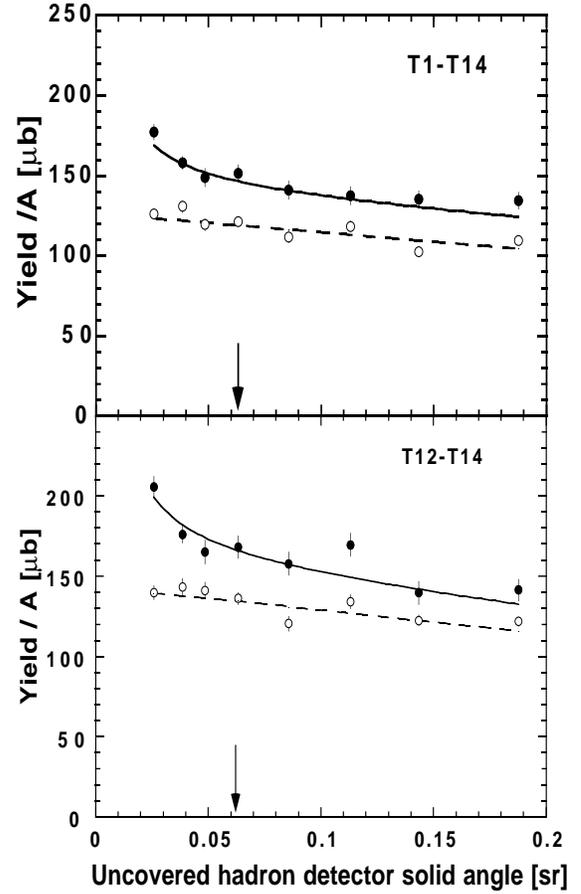}
\caption{The average yield on  all tagging channels (top) and on the three tagging channels at the lowest photon energies
(bottom) measured for different HD solid-angle coverage. 
The carbon data  for $E_0=2.8$ GeV (open circles) and the lead  data  for $E_0=2.2$ GeV (closed circles)
 are compared with the  MC predictions (dashed and solid lines respectively).
The arrow indicates the actual solid angle used for the measurement.}
\label{Fig5}
\end{figure}

The hadronic corrections, due to the finite angular acceptance of the HD and to the
possible contamination of electromagnetic events not vetoed by the SD, 
were experimentally tested by varying the HD solid angle coverage. This was performed by
moving  the target  upstream and downstream from
 the position used for the measurements.
The comparison between the MC and the experimental yields, for different HD solid angles,
 is shown in Fig.~\ref{Fig5}.
The average yields  for carbon and lead targets,
measured at $E_0$=2.8 GeV and  $E_0$=2.2 GeV respectively,
are given: the top panel refers to all tagging counters, while the bottom one to 
 the three tagging channels at the lowest photon energies.
In our geometry, the missing HD solid angle can be approximated as $\pi{\theta{^2}_{min}}$.
The MC predictions have been parameterized  in the form of 
$a - b\theta^{2}_{min} + c/ \theta^{2}_{min}$, where $a$ is the total cross section,   
$-b \theta^{2}_{min}$ represents the loss of 
hadronic events in the forward HD hole and 
$+c/ \theta^{2}_{min}$   represents the electromagnetic contamination 
 due to the $e^+e^-$ pairs produced in the target.
As shown in the figure, the hadronic losses are slowly increasing (yields decrease)
as the missing HD solid angle increases, while the electromagnetic contamination 
becomes relevant
 only at very small missing solid angle 
compared to the measurement position (yields increase). Moreover this contribution is  important only for the lead target 
and at the lowest photon energies. 
All the experimental yields are in good agreement with the MC predictions
 for both  contributions.

\begin{figure}[t]
\vspace{6cm}
\leavevmode
\includegraphics{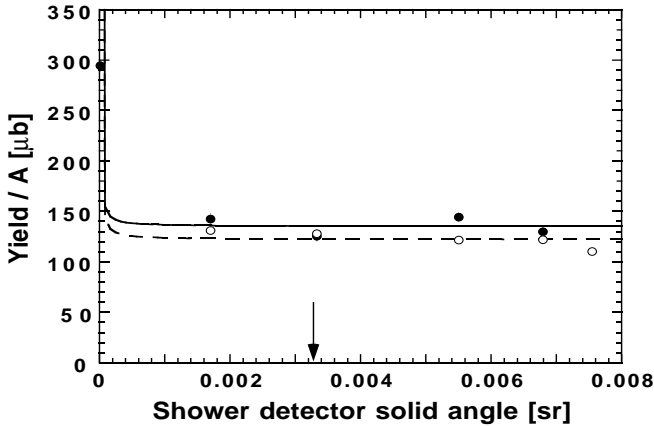}
\caption{The average C and Pb yields on all tagging channels, measured for different SD solid angle coverage.
 The notations are the same as for Fig.~\protect\ref{Fig5}.
 The arrow indicates the solid angle relevant for the 
measurement.}
\label{Fig6}
\end{figure}

The SD rejection efficiency as a function of the angular acceptance has been
evaluated by measuring the yields for varied radii of the C4 collimator.
These yields for carbon and lead, 
measured at $E_0$=2.8 GeV and  $E_0$=2.2 GeV respectively,
are shown in Fig.~\ref{Fig6} 
as a function of the forward solid angle $\Omega_{SD}  \approx  \pi \theta^2$ covered by the SD.
The arrow indicates the actual solid angle used for the measurements.
The MC predictions, parameterized  
in the form of $a + d/ \Omega_{SD}$, agree quite well with the experimental points.
As it is shown, the hadronic cross section is constant in a broad range of
  solid angle values, thus indicating that
  electromagnetic events were adequately suppressed by the veto counter.
An indication of the amount of the vetoed electromagnetic events
is given by the yield value at  $\Omega_{SD}$=0 sr, measured by removing the SD
veto.

Further experimental checks on the threshold efficiency of both
the HD and SD detectors have been performed, finding a very good agreement with the MC predictions.
 These checks are extensively described in Ref. \cite{MIR98}.

The results of the above described checks
validate the high reliability of the MC predictions. The average contribution of the
whole corrections, as a percent of the yields for all the studied nuclei
and at the three electron beam energies, are reported in Table~\ref{Table1}.
 
\vfill
\eject

\section{Results}

\subsection{Total cross sections}
\begin{figure}[t]
\vspace{6cm}
\leavevmode
\includegraphics{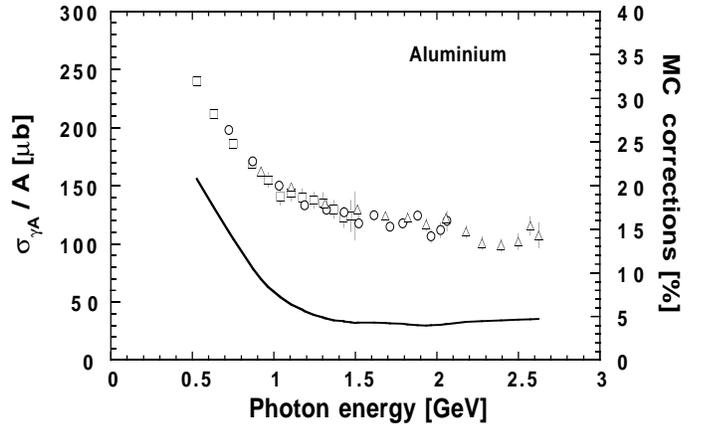}
\caption{Aluminium cross section measured at three electron beam energies: 1.6 GeV (squares),
2.2 GeV (circles), 2.8 GeV (triangles), on the left scale. MC correction (solid line)
 due to the hadronic losses 
and electromagnetic contaminations on the right scale.}
\label{Fig7}
\end{figure}

The cross section values were obtained by applying to the yields the previously described MC corrections.
As an example of the quality of the data in  Fig.~\ref{Fig7} 
 the cross section on aluminium,
measured at the three electron beam energies, is shown together with the MC corrections:
 the latter remain almost constant at about 5\%  in the region
of main interest for this measurement.
As to be seen the three data sets are well consistent within each
other.

The average values of the cross section 
in the overlapping regions
are given in Table~\ref{Table2}  for each nucleus.
These values  well agree among each other within the experimental errors
for all the studied nuclei, showing the good control of the systematic errors. 
 These mainly originate from the uncertainties
in the target thickness (reported in Table~\ref{Table1}),  the photon beam flux ($\approx$ 1\%),
the background subtraction ($\approx$ 1\% for C and $\approx$ 3\% for Pb), and the MC 
corrections ($\approx$ 1.5\% for C and $\approx$ 2.5\% for Pb). The  total 
average systematic errors are $\approx$ 2\% for C and $\approx$ 5\% for Pb.
 
\begin{table}[thb]
\caption{Average cross section [$\mu$b$/A$] in the overlapping photon energy regions $\Delta k$ [MeV] for the three
 beam energies $E_0$ [GeV] and for each target T.
The errors are the quadratic sum of statistical and systematic uncertainties. }
\begin{center}
\begin{tabular}{lc@{\hspace{0.2cm}}c@{\hspace{0.2cm}}c@{\hspace{0.2cm}}c@{\hspace{0.2cm}}c@{\hspace{0.2cm}}c}\hline \hline
        & \multicolumn{2}{c}{$\Delta k$=690-1480 }
        & \multicolumn{2}{c}{$\Delta k$=830-1990 }
        & \multicolumn{2}{c}{$\Delta k$=830-1480 }\\
\cline{2-7}
        & $E_0$ & $E_0$ & $E_0$ & $E_0$ & $E_0$ & $E_0$ \\
T & 1.6       &   2.2     & 2.2      & 2.8       & 1.6      & 2.8 \\ \hline
 C      & 156$\pm$6 &   159$\pm$4 & 137$\pm$4 & 142$\pm$4 & 151$\pm$ 7 & 157$\pm$8 \\
 Al     & 158$\pm$8 &   159$\pm$4 & 138$\pm$5 & 142$\pm$4 & 150$\pm$ 8 & 153$\pm$4 \\
 Cu     & 165$\pm$10 &  163$\pm$5 & 143$\pm$6 & 147$\pm$5 & 158$\pm$11 & 157$\pm$5 \\
 Sn     & 167$\pm$11 &  164$\pm$6 & 142$\pm$6 & 144$\pm$6 & 160$\pm$11 & 155$\pm$6 \\
 Pb     & 154$\pm$12 &  153$\pm$8 & 137$\pm$8 & 144$\pm$8 & 149$\pm$13 & 147$\pm$7 \\
\hline \hline
\end{tabular}
\label{Table2}
\end{center}
\end{table}
      
The cross section data measured at the three beam energies have been partitioned and averaged in 19 
bins of energy about 100 MeV wide. The resulting total cross section values, normalized to the mass number A,
are given in Table~\ref{Table3} for all the studied nuclei together with the statistical errors.

\begin{figure}[ht]
\vspace{15.8cm}
\leavevmode
\includegraphics{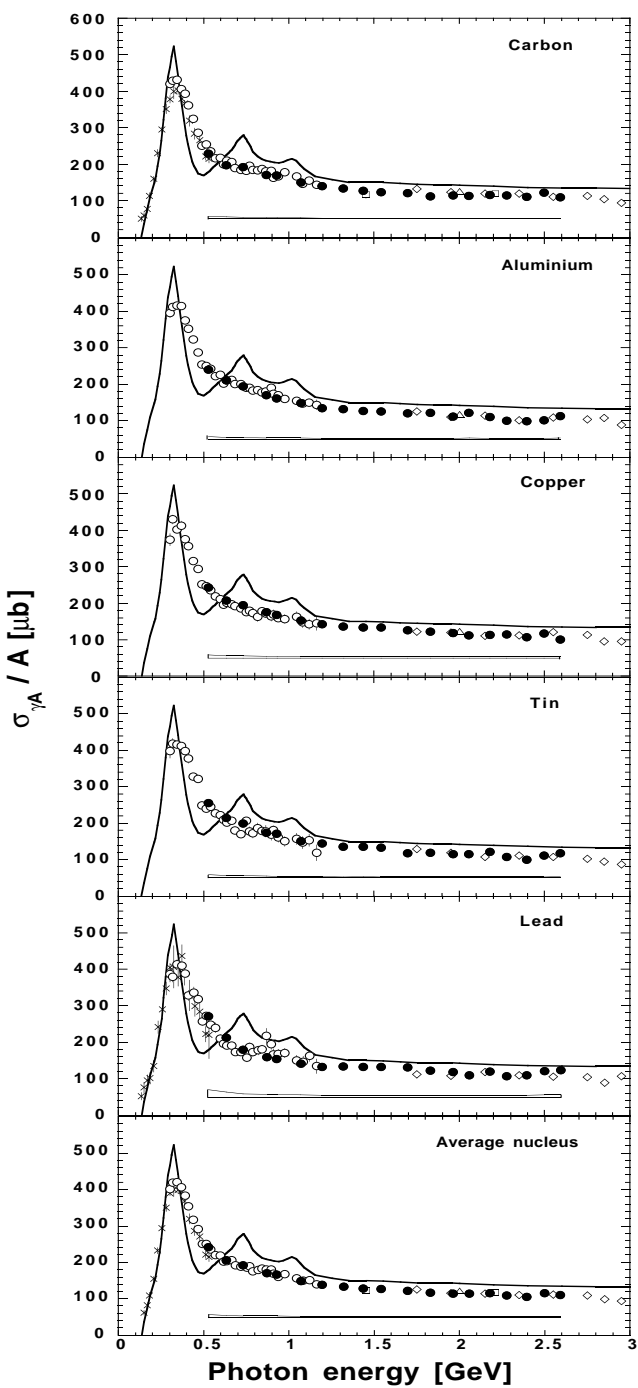}
\caption{Total cross section data on C, Al, Cu, Sn, Pb and average nucleus (solid circles)
compared with data from previous experiments: crosses \protect\cite{CAR85},
 open circles \protect\cite{BIA96},
 squares \protect\cite{HEY71}, diamonds 
\protect\cite{BRO73}, and triangles 
\protect\cite{MIC77}. 
Also shown is the proton 
absorption cross section \protect\cite{PDG} (solid line).
The widths of the bands represent the systematic errors.}
\label{Fig8}
\end{figure}

In the last column  the weighted average of cross sections for the five nuclei is also given.  This can be considered
as the cross section on an average nucleus with Z/A = 0.469 and an average nuclear density
$\rho_{A}$=0.109 nucleons/fm$^{3}$.
The photoabsorption cross sections  are also shown in Fig.~\ref{Fig8}, together with the data
for the proton \cite{PDG}.

The bars indicate the statistical errors only,  the
band in the bottom of the panels represent the systematic uncertainties. 
The present data are in very good agreement within the experimental errors with both the
low and high energy data available in the literature.
They confirm with reduced statistical uncertainties the absence of peaks in the
region of the second and third resonances for
the bound nucleon.
The new and  most striking result is the persistence of the absorption strength  reduction
 above 1.2 GeV compared to the free nucleon case for all the studied nuclei. 

\begin{table}[h]
\caption{Total cross sections [$\mu$b] and statistical errors normalized to the mass
 number A at the photon energy $k$ [GeV].
The average value $\overline{A}$ is calculated weighting each nucleus cross section value with its
statistical error. }
\begin{center}
\begin{tabular}{cc@{\hspace{0.2cm}}c@{\hspace{0.2cm}}c@{\hspace{0.2cm}}c@{\hspace{0.2cm}}c@{\hspace{0.2cm}}c} \hline \hline
     & \multicolumn{6}{c}{Total\, cross\, section  /A} \\   
 $k$    &  C        & Al        & Cu        &    Sn     & Pb        & $ \overline{A}$ \\ \hline
0.53 & 229$\pm$3 & 240$\pm$3 & 244$\pm$4 & 256$\pm$4 & 271$\pm$4 & 243$\pm$2 \\
0.63 & 197$\pm$3 & 211$\pm$3 & 208$\pm$4 & 215$\pm$4 & 213$\pm$5 & 207$\pm$2 \\
0.73 & 192$\pm$1 & 195$\pm$2 & 195$\pm$3 & 199$\pm$2 & 180$\pm$3 & 193$\pm$1 \\
0.87 & 170$\pm$1 & 170$\pm$2 & 176$\pm$2 & 174$\pm$3 & 160$\pm$3 & 171$\pm$1 \\
0.93 & 169$\pm$2 & 161$\pm$3 & 169$\pm$2 & 171$\pm$3 & 154$\pm$3 & 167$\pm$1 \\
1.07 & 150$\pm$1 & 148$\pm$2 & 153$\pm$2 & 152$\pm$2 & 142$\pm$2 & 150$\pm$1 \\
1.19 & 139$\pm$2 & 134$\pm$3 & 142$\pm$3 & 145$\pm$3 & 134$\pm$3 & 139$\pm$1 \\
1.32 & 134$\pm$1 & 132$\pm$2 & 136$\pm$2 & 136$\pm$3 & 133$\pm$3 & 134$\pm$1 \\
1.43 & 126$\pm$2 & 126$\pm$3 & 134$\pm$4 & 137$\pm$5 & 133$\pm$5 & 129$\pm$2 \\
1.54 & 123$\pm$1 & 125$\pm$2 & 134$\pm$2 & 134$\pm$3 & 132$\pm$3 & 127$\pm$1 \\
1.70 & 121$\pm$2 & 120$\pm$3 & 126$\pm$3 & 118$\pm$4 & 130$\pm$4 & 122$\pm$1 \\
1.83 & 112$\pm$2 & 121$\pm$3 & 122$\pm$3 & 119$\pm$4 & 122$\pm$4 & 117$\pm$1 \\
1.96 & 114$\pm$2 & 112$\pm$3 & 118$\pm$4 & 116$\pm$4 & 119$\pm$5 & 115$\pm$1 \\
2.06 & 114$\pm$3 & 121$\pm$5 & 112$\pm$5 & 116$\pm$6 & 110$\pm$6 & 115$\pm$2 \\
2.18 & 116$\pm$4 & 110$\pm$5 & 114$\pm$5 & 122$\pm$6 & 119$\pm$7 & 116$\pm$2 \\
2.28 & 114$\pm$4 & 100$\pm$6 & 115$\pm$6 & 108$\pm$7 & 107$\pm$8 & 110$\pm$3 \\
2.39 & 111$\pm$4 & 98 $\pm$5 & 107$\pm$6 & 101$\pm$7 & 109$\pm$8 & 106$\pm$2 \\
2.50 & 122$\pm$5 & 102$\pm$7 & 117$\pm$7 & 112$\pm$8 & 122$\pm$9 & 116$\pm$3 \\
2.59 & 109$\pm$5 & 112$\pm$7 & 101$\pm$7 & 118$\pm$9 & 124$\pm$10& 111$\pm$3 \\
\hline \hline
\end{tabular}
\label{Table3}
\end{center}
\end{table}

\subsection{Photonuclear to photonucleon cross section ratio}
In order to better evaluate this strength reduction, the ratio between the nuclear
cross section $\sigma_{\gamma A}$ and that obtained for the free nucleons
 (Z$\sigma_{\gamma p}$+N$\sigma_{\gamma n}$),
 derived from proton \cite{PDG} and deuteron data\cite{BIA96,BAB98}, has been calculated in each energy region.
These ratios are shown in Fig.~\ref{Fig9}, together with the results
 of a $\Delta$-hole model  \cite{CA92} and  
of two recent VMD calculations for C, Cu and Pb \cite{PI95,BO96}.

The former calculation well reproduces the experimental behavior at
lower energies, while both VMD calculations  do not
predict the systematic nuclear damping of the cross section clearly indicated by this experiment.
In addition, 
 the inclusion of two-nucleon correlations considered in Ref. \cite{BO96}
 leads to an anti-shadowing behavior below 2 GeV and thus to an even larger disagreement with the data.


\subsection{Photoabsorption strength in the nuclear medium}

 For each nucleus the  strength reduction was evaluated   
 in the five 
energy regions of mean energy $\overline{k}$ given in Table~\ref{Table4}.
Both the 
integral $\Sigma_A (\overline{k})$ of the measured cross sections and
 the ratio $R_A (\overline{k})={\Sigma_A}/({Z\Sigma_p + N \Sigma_n })$ were calculated. 
 Here
$\Sigma_{p}$ and $\Sigma_{n}$ are the proton and neutron cross sections integrated
over the relevant photon energies.

In Table~\ref{Table4} the averaged $\overline{R}_{A}(\overline{k})$,
computed weighting $R_{A}(\overline{k})$ for each nucleus with its statistical error,
 is given.
The energy behavior of $\overline{R}_{A}(\overline{k})$  is compared in Fig.~\ref{Fig10}a)
 with the one derived from data from previous experiments.
 The comparison evidences the good agreement between data in both the resonance and the 
shadowing regions.
In the shadowing threshold region data from the present experiment evidence a photoabasorption strength reduction
well above the experimental errors.

In addition the nuclear density dependence of  ${\Sigma_{A}(\overline{k})}$ in each region
was parameterized in the form 
\begin{equation} 
        \frac{\Sigma_A (\overline{k})}{A} = \Sigma_0(\overline{k})[1+\beta(\overline{k})\rho_A], 
\end{equation}
where 
 $\rho_A$ is the average nuclear density. The latter was derived from the experimental charge density distributions 
 with the rms electron-scattering radius $\overline{r^2}$ 
given in Ref. \cite{DEJ74}.
Different parameterizations of the nuclear density distribution result in an average variation of less than 5$\%$ 
 in the  $\rho_A$ value.
\begin{table}[t]
\caption{Energy regions, corresponding ranges and mean energies $\overline{k}$ [GeV], and average 
ratios $\overline{R}_A (\overline{k})$. The latter were
computed by weighting $R_A(\overline{k})$ for each nucleus with its statistical error.
In the last column, also the statistical and systematic errors are reported. }
\begin{center}
\begin{tabular}{c@{\hspace{0.2cm}}c@{\hspace{0.2cm}}c@{\hspace{0.2cm}}c} \hline \hline 
\\
Energy region            & Range   & $\overline{k}$ & $\overline{R}_A (\overline{k})$\\                       
\hline
$\Delta$-resonance tail  &0.48-0.68   &  0.58    & 1.170$\pm$0.009$\pm$0.043  \\
$ D_{13}$-resonance      &0.68-0.88   &  0.80    & 0.771$\pm$0.004$\pm$0.021  \\      
$ F_{15}$-resonance      &0.88-1.25   &  1.00    & 0.936$\pm$0.006$\pm$0.024  \\
shadowing threshold      &1.25-1.65   &  1.40    & 0.847$\pm$0.007$\pm$0.030  \\
shadowing                &1.65-2.65   &  2.00    & 0.863$\pm$0.015$\pm$0.023  \\
\hline \hline
\end{tabular}
\label{Table4}
\end{center}
\end{table}

In Fig.~\ref{Fig10}b) 
 the coefficients  $\beta$ obtained in each energy region from the fits to our data are shown.
Also shown are the  $\beta$ derived from  data from previous photoabsorption experiments.
The energy behavior of  $\beta$  is similar to the one of $\overline{R}_{A}$
  in the resonance region
 and above 3 GeV.
This indicates that the mechanisms responsible 
for the observed medium effects depend on the nuclear density.

On the contrary, in the shadowing threshold region the $\beta$ is positive  while
 $\overline{R}_{A}$ is less than unity,
indicating a  stronger  strength reduction in the light nuclei.

Similar information can be obtained also from different reactions, specifically from electron scattering at
low momentum transfer $Q^2$ and from  photon scattering at  small-angle.
Then the $\beta$  have  been derived from  measurements of these reactions
 performed on a wide range of mass number  \cite{BAY79},  \cite{CRI77}.
In the framework of  a VMD  description  the relevant quantity is 
the coherence length of the hadronic fluctuations of the photon
$\lambda_V= 2k /(Q^2 + m_V^2)$ in the laboratory frame, where $ m_V$ is the vector-meson mass.
 At low energy the hadronic fluctuations of the photon are dominated by the  $\rho$-meson.

\vspace{22cm}

Therefore in Fig.~\ref{Fig10bis} the $\beta$ values are shown as a function of $\lambda_{\rho}$ 
which is the coherence length when only the
 $\rho$-meson contribution is considered. The overall good agreement
 points out the consistency of total photoabsorption, electron scattering and photon scattering data.
In particular, in the shadowing 
threshold region ($\lambda_{\rho}<1.3$ fm) this agreement strengthens 
the evidence of a larger medium effect in
light nuclei.
This experimental finding may suggest in this region a different mechanism for the 
strength reduction, which does not depend on the nuclear density alone.

\vspace{22cm}
\begin{figure}[ht]
\vspace{15.8cm}
\leavevmode
\includegraphics{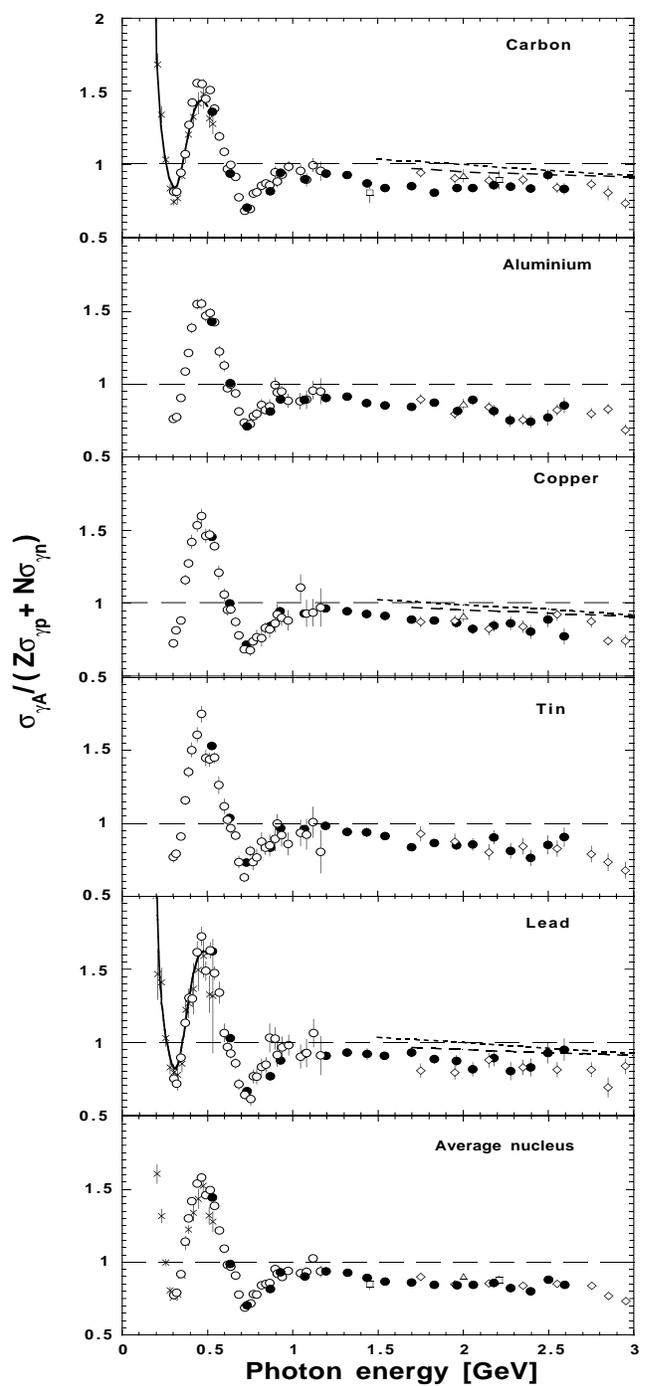}
\caption{Ratio of photonuclear and photonucleon absorption cross sections. Same notation as for Fig.~\protect\ref{Fig8}.
Solid line is a ${\Delta}$-hole model \protect\cite{CA92}, dashed   \protect\cite{PI95}
and dotted  \protect\cite{BO96} lines are VMD predictions. The bars represent the statistical errors.}
\label{Fig9} 
\end{figure}

This  could be due to the shadowing onset at lower energy for ligth nuclei.
The shadowing effect at higher energy is 
generally described by VMD models.
These models,  which consider the photon as a superposition of a bare photon and 
vector mesons with $\delta$-function mass distribution, 
are able to reproduce the photo-nuclear absorption cross section
 in the several GeV domain \cite{BA78, PI95, BO96} and 
shadowing phenomena observed in deep-inelastic lepton-nucleus scattering in the low
x region (x being the Bjorken variable) \cite{DI76, SH93, PI90}.

\vspace{10cm}
\begin{figure}[ht]
\vspace{10cm}
\leavevmode
\includegraphics{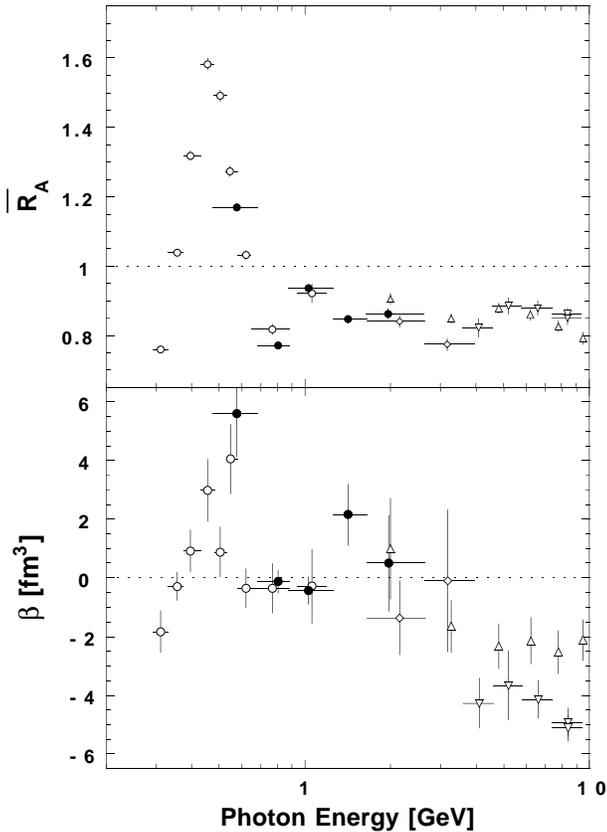}
\caption{a) Average ratio $\overline{R}_{A}$ and b) linear coefficient $\beta$ derived from our data 
(solid circles), 
and from Refs. \protect\cite{BIA96} (open circles), \protect\cite{BRO73} (diamonds),
 \protect\cite{MIC77} (upper triangles), and
\protect\cite{CAL73} (lower triangles). The vertical bars represent the statistical errors, while
the horizontal bars represent the bin widths.}
\label{Fig10}
\end{figure}

However, as shown in Fig.~\ref{Fig9}, the two more recent
VMD calculations for real photons \cite{PI95,BO96}
 clearly
 underestimate the shadowing effect below 3 GeV. Therefore, 
a  model which takes into account a more realistic 
 spectral function of the low mass hadronic components should be considered in order to
better reproduce the photoabsorption data.

In addition, 
a possible change of the vector-meson properties in the nuclear medium 
(a reduction of the vector-meson mass
 \cite{BR91,HAT92} 
  and the modifications of the $\rho$-meson spectral function
 \cite{RAP97,PET98}) can  also be considered. This 
 enanches the strenght at small invariant 
mass, thus contributing to reduce the energy threshold for the shadowing effect.



\begin{figure} [ht]
\vspace{6cm}
\leavevmode
\includegraphics{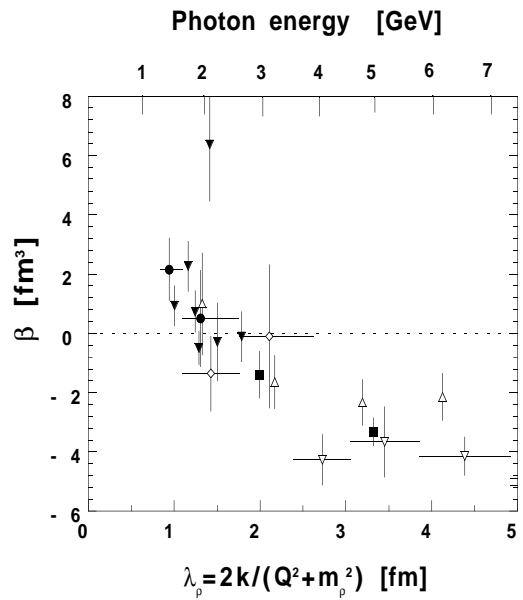}
\caption{Linear coefficient $\beta$ derived from photoabsorption data (same notation as for 
Fig.~\protect\ref{Fig10}), electron scattering \protect\cite{BAY79} (solid triangles) and 
photon scattering  
\protect\cite{CRI77} (solid squares) data. The vertical bars represent the statistical errors, while
the horizontal bars represent the bin widths.}
\label{Fig10bis}
\end{figure}

\section{Conclusions}

The total photoabsorption cross section on C, Al, Cu, Sn and Pb has been measured  in the energy range 0.5-2.6 GeV.

 The data confirme the absence of structures in the $D_{13}$ and $F_{15}$ resonance region
 and show the damping of the photoabsorption
strength
above 0.6 GeV compared to the free nucleon case.
The new result is the persistence of the strength reduction in the unexplored energy region 1.2-1.7 GeV, where resonance effects
are expected to be small. 
In addition, our systematic measurement over a wide range of mass numbers indicates that 
in this region the strength
reduction decreases  with the nuclear density.

This reduction can be interpreted as a signature of a low energy onset of the shadowing effect.

In the framework of a VMD description of the shadowing effect 
both a decrease of the vector meson mass and a significant broadening of the $\rho$-meson spectral function 
in the nuclear medium, can produce an earlier onset of the shadowing effect.
In this respect the new photoabsorption data provide interesting insigths concerning the 
modification of the vector meson properties in the nuclear medium.

\section{Acknowledgements}

We would like to express our gratitude to the Frascati technicians A. Orlandi, W. Pesci, G.F Serafini and 
A. Viticchi\'e for their continuous technical assistance. We are pleased to thank the ELSA staff and the  SHAPIR group for the
 efficiency in running the machine and the tagged photon beam.


\end{document}